\documentclass[a4paper]{jpconf}
\usepackage{graphicx}

\usepackage{graphicx}
\usepackage{hyperref}

\graphicspath{{./fig/}}

\begin{document}
\title{Tunka Advanced Instrument for cosmic rays and Gamma Astronomy}

\author{
D~Kostunin$^{4,15}$,
I~Astapov$^{1}$,
P~Bezyazeekov$^{2}$,
A~Borodin$^{3}$,
N~Budnev$^{2}$,
M~Br\"uckner$^{4}$,
A~Chiavassa$^{5}$,
A~Dyachok$^{2}$,
O~Fedorov$^{2}$,
A~Gafarov$^{2}$,
A~Garmash$^{6}$,
V~Grebenyuk$^{3,7}$,
O~Gress$^{2}$,
T~Gress$^{2}$,
O~Grishin$^{2}$,
A~Grinyuk$^{3}$,
A~Haungs$^{15}$,
D~Horns$^{8}$,
T~Huege$^{15,16}$,
A~Ivanova$^{2}$,
N~Kalmykov$^{9}$,
Y~Kazarina$^{2}$,
V~Kindin$^{1}$,
P~Kirilenko$^{6}$,
S~Kiryuhin$^{2}$,
M~Kleifges$^{17}$,
R~Kokoulin$^{1}$,
K~Kompaniets$^{1}$,
E~Korosteleva$^{9}$,
V~Kozhin$^{9}$,
E~Kravchenko$^{6,10}$,
A~Kryukov$^{9}$
L~Kuzmichev$^{9,2}$,
Yu~Lemeshev$^{2}$,
A~Lagutin$^{19}$,
V~Lenok$^{15}$,
B~Lubsandorzhiev$^{11}$,
N~Lubsandorzhiev$^{9}$,
T~Marshalkina$^{2}$,
R~Mirgazov$^{2}$,
R~Mirzoyan$^{12,2}$,
R~Monkhoev$^{2}$,
E~Osipova$^{2}$,
A~Pakhorukov$^{2}$,
A~Pan$^{3}$,
M~Panasyuk$^{9}$,
L~Pankov$^{2}$,
A~Petrukhin$^{1}$,
V~Poleschuk$^{2}$,
M~Popescu$^{13}$,
E~Popova$^{9}$,
A~Porelli$^{4}$,
E~Postnikov$^{9}$,
V~Prosin$^{9}$,
V~Ptuskin$^{14}$,
A~Pushnin$^{2}$,
R~Raikin$^{19}$,
E~Ryabov$^{2}$,
G~Rubtsov$^{11}$,
Y~Sagan$^{3,7}$,
B~Sabirov$^{3}$,
V~Samoliga$^{2}$,
Yu~Semeney$^{2}$,
F~G~Schr\"oder$^{15,18}$,
A~Silaev$^{9}$,
A~Silaev(junior)$^{9}$,
A~Sidorenkov$^{11}$,
A~Skurikhin$^{9}$,
V~Slunecka$^{3}$,
A~Sokolov$^{6,10}$,
C~Spiering$^{4}$,
L~Sveshnikova$^{9}$,
V~Tabolenko$^{2}$,
B~Tarashansky$^{2}$,
L~Tkachev$^{3,7}$,
M~Tluczykont$^{8}$,
N~Ushakov$^{11}$,
A~Vaidyanathan$^{6,10}$
P~Volchugov$^{9}$,
D~Voronin$^{11}$,
R~Wischnewski$^{4}$,
A~Zagorodnikov$^{2}$,
D~Zhurov$^{2}$,
V~Zurbanov$^{2}$ and
I~Yashin$^{1}$
}

\address{$^{1}$ National Research Nuclear University MEPhI (Moscow Engineering Physics Institute), Moscow, Russia}
\address{$^{2}$ Institute of Applied Physics ISU, Irkutsk, Russia}
\address{$^{3}$ JINR, Dubna, Russia}
\address{$^{4}$ DESY, Zeuthen, Germany}
\address{$^{5}$ Dipartimento di Fisica Generale Universiteta di Torino and INFN, Torino, Italy}
\address{$^{6}$ Novosibirsk State University, NSU, Novosibirsk, Russia}
\address{$^{7}$ Dubna State University, Dubna, Russia}
\address{$^{8}$ Institute for Experimental Physics, University of Hamburg, Germany}
\address{$^{9}$ Skobeltsyn Institute of Nuclear Physics MSU, Moscow, Russia}
\address{$^{10}$ Budker Institute of Nuclear Physics SB RAS, Novosibirsk, Russia}
\address{$^{11}$ Institute for Nuclear Research of RAN, Moscow, Russia}
\address{$^{12}$ Max-Planck-Institute for Physics, Munich, Germany}
\address{$^{13}$ ISS, Bucharest, Romania}
\address{$^{14}$ IZMIRAN, Moscow, Russia}
\address{$^{15}$ Institut f\"ur Kernphysik, Karlsruhe Institute of Technology (KIT), Karlsruhe, Germany}
\address{$^{16}$ Astrophysical Institute, Vrije Universiteit Brussel, Brussels, Belgium}
\address{$^{17}$ Institut f\"ur Prozessdatenverarbeitung und Elektronik, Karlsruhe Institute of Technology (KIT), Karlsruhe, Germany}
\address{$^{18}$ Bartol Research Inst., Dept. of Phys. and Astron., Univ. of Delaware, Newark, DE, USA}
\address{$^{19}$ Altai State University, Barnaul, Russia}

\ead{dmitriy.kostunin@desy.de}

\begin{abstract}
The paper is a script of a lecture given at the ISAPP-Baikal summer school in 2018.
The lecture gives an overview of the Tunka Advanced Instrument for cosmic rays and Gamma Astronomy (TAIGA) facility including historical introduction, description of existing and future setups, and outreach and open data activities.
\end{abstract}

\section{Introduction}

The nature of the most energetic phenomena of the Universe is still under cover of mystery.
They happen farer than tens to hundreds light years from the Earth (luckily for the life on the planet) and we receive very fragmented information from such distances.
For the time being modern science has discovered all known messengers connected to the fundamental interactions: photons (electromagnetism), cosmic rays (strong force), neutrinos 
(weak force) and gravitational waves (gravitation).
In this work we will discuss two of these messengers, namely the very- and ultra-high energy cosmic and gamma rays, what kind of information they carry and see how they can be detected, with a focus on the instrumentation installed in the Tunka Valley in Russia, near the southern tip of Lake Baikal.

Cosmic rays are nuclei ranging from hydrogen to iron, which are produced and accelerated in non-thermal processes in the Universe, e.g. in supernova explosions, ultra-relativistic jets from compact objects, etc.~\cite{Ptitsyna:2008zs, Troitsky:2013rxa}.
By measuring the flux and composition of them we can decipher the processes happening in the accelerator.
Particularly it was found that the transition from galactic to extragalactic accelerators happens in the energy range between PeV and EeV, i.e. the most energetic particles are accelerated outside of our Galaxy.
Unfortunately, due to their charge, cosmic rays are heavily declined in cosmic magnetic fields, what makes it impossible to trace them back to the source.
Modern detectors can resolve only large-scale anisotropies in their direction~\cite{Ahlers:2016njd,Aab:2017tyv}, which gives only hints regarding the sources. 
Contrary to charged cosmic rays, neutral gamma rays of energies in the TeV-PeV region, which are generated in the processes strongly connected to the production of cosmic rays, can directly point to the source of their production~\cite{Aharonian:2006au}.

The flux of these high-energetic particles falls steeply (with power of $\gamma \approx-2.7$), reaching one particle per square kilometer in thousand years, which at the moment makes their direct detection impossible.
Nevertheless they can be detected after collision with the Earth's atmosphere, which produces a particle cascade radiating at different wavelengths and expanding to tens of square kilometers at the ground level.
These cascades, which are called extensive air showers, can be either detected by sparse ground arrays measuring secondary particles and radiation, or by Cherenkov and fluorescence telescopes measuring light directly in the atmosphere.

\begin{figure}[t]
\includegraphics[width=1.0\linewidth]{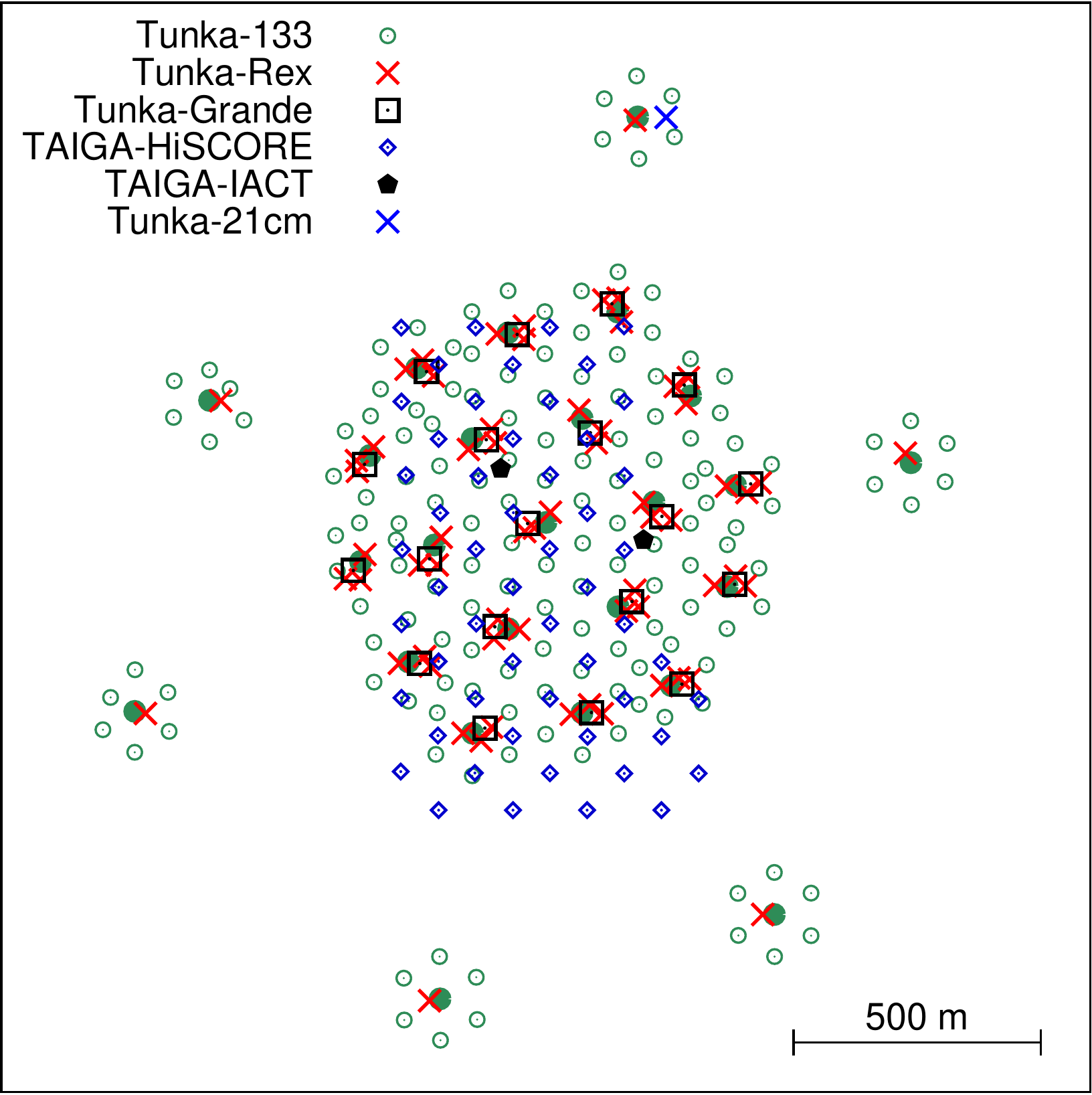}
\caption{Layout of TAIGA observatory. Cosmic-ray setups are arranged in 25 clusters, solid circles indicate cluster centers.}
\label{fig:taiga-map}
\end{figure}

One of these arrays is installed in the Tunka Valley in Eastern Siberia near the southern tip of Lake Baikal.
The Tunka Advanced Instrument for cosmic rays and Gamma Astronomy
(TAIGA)~\cite{Budnev:2018fxf} is a facility equipped with three detectors
measuring secondary particles, air-Cherenkov light and radio emission from
air-showers created by primary cosmic rays with energies in the PeV-EeV region
as well as with telescopes and a timing array measuring air-Cherenkov
light created by gamma-ray air showers with energies of TeV-PeV.
Combining the sensitivity of all setups, TAIGA covers four orders of magnitude in the cosmic-ray energy spectrum, overlapping satellite measurements at the lower energies and the transition from galactic to extragalactic components at the highest ones.
For the time being TAIGA covers an area of 3 km\textsuperscript{2}, and is planned to be extended and equipped with new detectors.
One can see the layout of the facility in Fig.~\ref{fig:taiga-map}, the particular setups will be described below. 

Besides the main activity of TAIGA related to astroparticle physics, we will discuss the cooperation with experiments located in the same area (Tunka) as well as the support of education, outreach and open data policies, and give a short historical introduction.

\section{The history of the Tunka astrophysical facility}

\begin{figure}[t]
\includegraphics[width=1.0\linewidth]{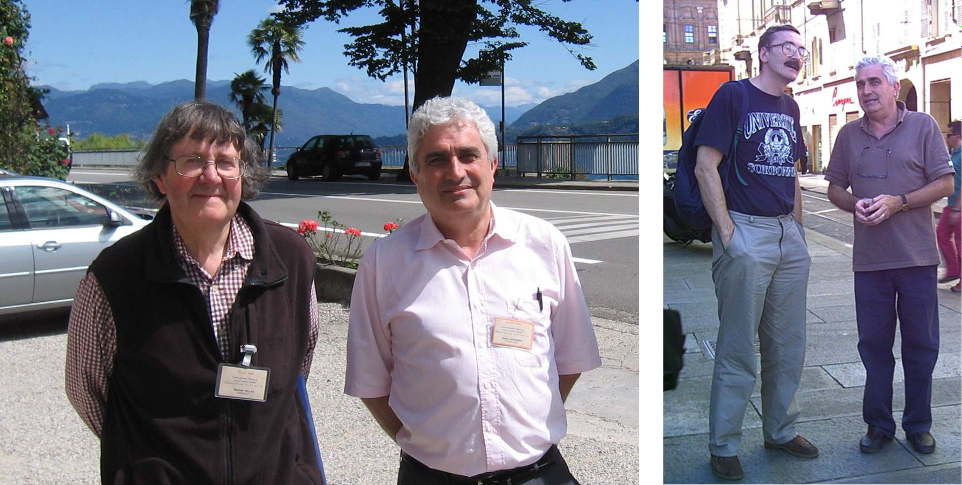}
\caption{\textit{Left:} A.~M.~Hillas (1932--2017) and G.~Navarra (1945--2009) at the International Cosmic Ray Conference in Rome in 1995. \textit{Right:} Leonid Kuzmichev, the head of the Tunka facility, is discussing the Tunka-133 project with the late Gianni Navarra         (spokesperson of EAS-TOP) in Torino.}
\label{fig:hillas_navarra_kuzmichev}
\end{figure}

\begin{figure}[t]
\begin{center}
\includegraphics[width=1.0\textwidth]{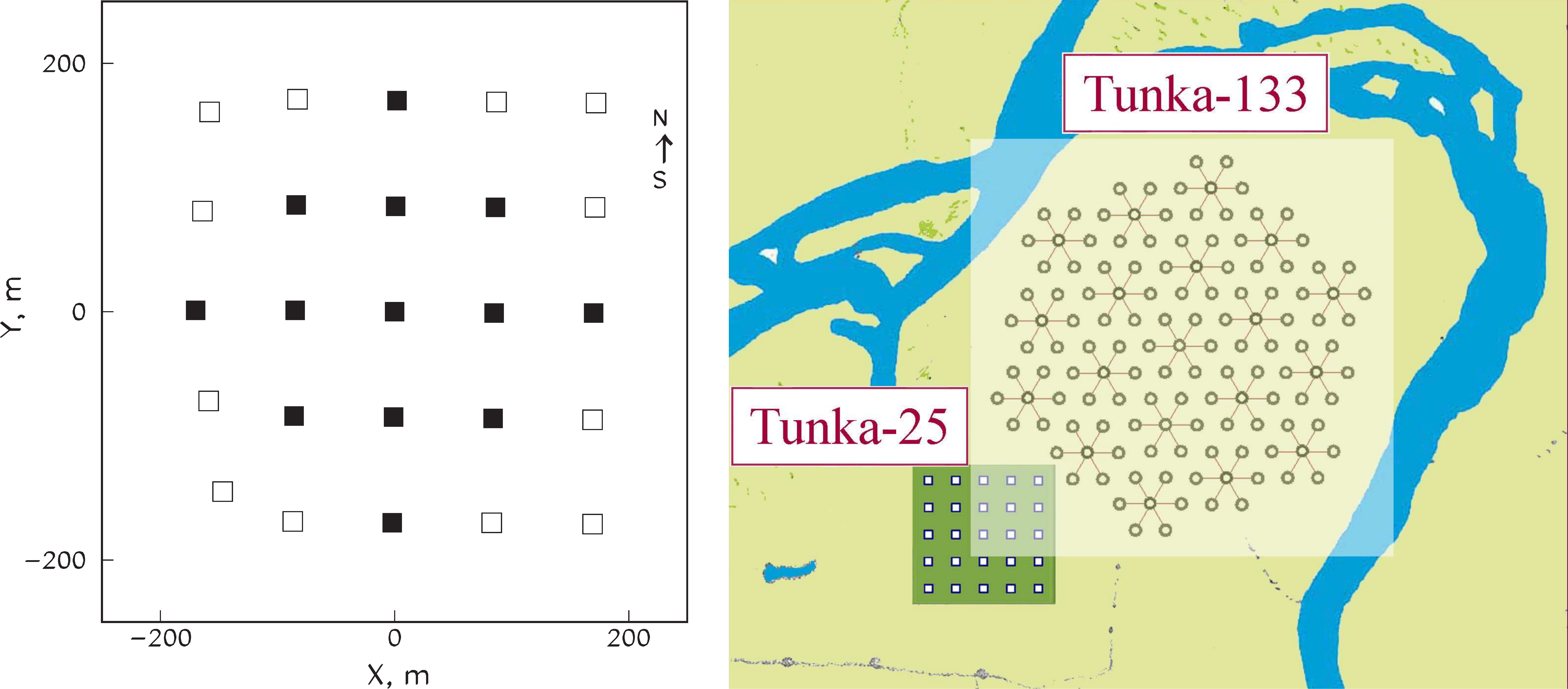}
\caption[Development stages of the Tunka-133 detector]{Development stages of the Tunka-133 air-Cherenkov detector.
\textit{Left:}~Tunka-25 detector~\cite{Budnev201318}. Black points are the original Tunka-13 detector~\cite{Gress1999299} triggering Tunka-25.
\textit{Right:}~First stage of the Tunka-133 detector~\cite{Antokhonov:2011zza} of the year 2009. Six additional satellite clusters were deployed in 2011.}
\label{fig_tunka25_133}
\end{center}
\end{figure}

The predecessor of the astrophysical activity in the Tunka Valley was a test setup with four PMTs QUASAR-370 (manufactured in the Soviet Union) on the ice of Lake Baikal.
Later this pathfinder array was relocated to the valley and formed the first engineering array Tunka-4.
After the ICRC conference in 1995, the Tunka experiment received support from Gianni~Navarra and Andrew~Michael~Hillas (see some historical photos in~Fig.~\ref{fig:hillas_navarra_kuzmichev}).
This collaboration led to the calibration of the forthcoming air-Cherenkov detector Tunka-13~\cite{Gress1999299} with the QUEST experiment~\cite{Korosteleva:2007ek} co-located with EAS-TOP at LNGS.
Later, the array was extended to Tunka-25~\cite{Budnev201318} covering a square area of 0.1~km\textsuperscript{2} (see~Fig.~\ref{fig_tunka25_133}).
In the beginning of the 2000s, the PMTs from the former MACRO experiment were shipped to Tunka, and deployed in the optical modules of the Tunka-133 detector covering 1 km\textsuperscript{2}~\cite{Antokhonov:2011zza}, which was later extended to 3~km\textsuperscript{2} by installing six satellite clusters.

In 2012 in the frame of Russian-German joint research activities, the construction of two new detectors was started.
The first one, a digital radio array called Tunka Radio Extension (Tunka-Rex)~\cite{Bezyazeekov:2015rpa}, was deployed and commissioned in the same year.
The second one, a low-threshold non-imaging air-Cherenkov array originally called Tunka-HiSCORE (Hundred*i Square-km Cosmic ORigin Explorer) and later TAIGA-HiSCORE (High-Sensitivity Cosmic ORigin Explorer) was put into operation one year later, in 2013~\cite{Tluczykont:2014cva}. The deployment of the array has started the re-orientation of the Tunka facility towards gamma detection, and the entire facility was later called TAIGA (Tunka Advanced Instrument for cosmic rays and Gamma Astronomy).

In 2014-2015 in the frame of an agreement between MSU and KIT, the former KASCADE-Grande scintillators were shipped to Tunka and formed the Tunka-Grande array~\cite{Monkhoev:2017yhj}, which could deliver a full-day trigger for Tunka-Rex.
At the same time the TAIGA-HiSCORE and Tunka-Rex detectors have been several times extended. Tunka-Rex reached its current layout in 2016.

Meanwhile the construction of a net of imaging atmospheric Cherenkov telescopes (IACT) was granted by the Russian government, and the first telescope, TAIGA-IACT was deployed in 2017~\cite{Lubsandorzhiev:2017tby, Zhurov:2017uvs}.
The second telescope of the net was constructed one year later, in 2018.

Since 2018 TAIGA is supporting open data policies in the frame of the German-Russian Astroparticle Data Life Cycle (GRADLC) Consortium~\cite{Bychkov:2018zre}.
The part of software and data measured by the TAIGA experiments (mostly by Tunka-Rex) are open, the next release is planned to be done in cooperation with KCDC project~\cite{Haungs:2018xpw}.
Having unique experience in constructing and maintaining cosmic-ray setups as well as in data analysis, the members of TAIGA collaboration will share their knowledge in the future educational and outreach program of GRADLC.

\section{Cosmic-ray setups}

\begin{figure}[t]
\includegraphics[width=1.0\linewidth]{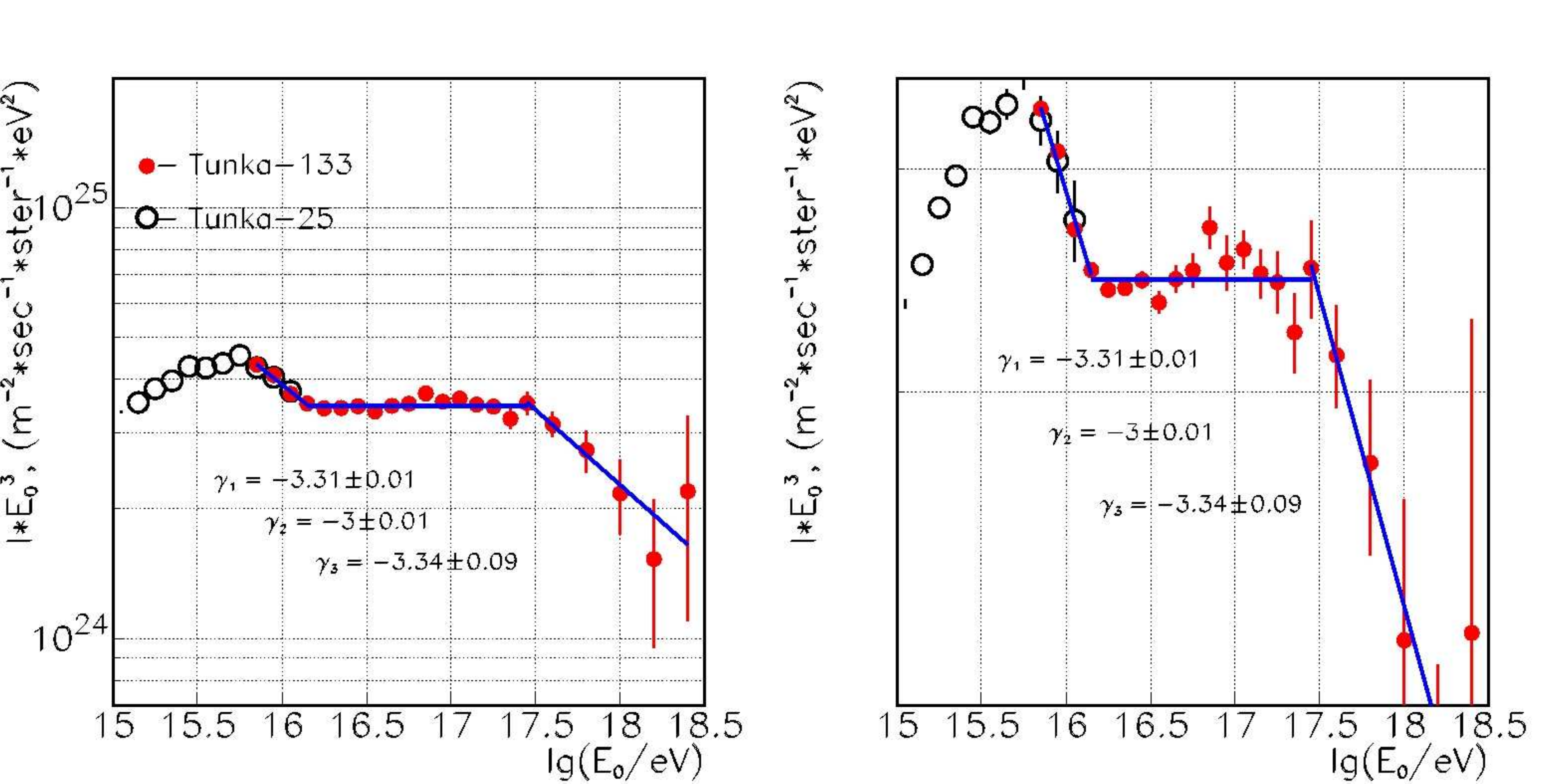}\\
\includegraphics[width=1.0\linewidth]{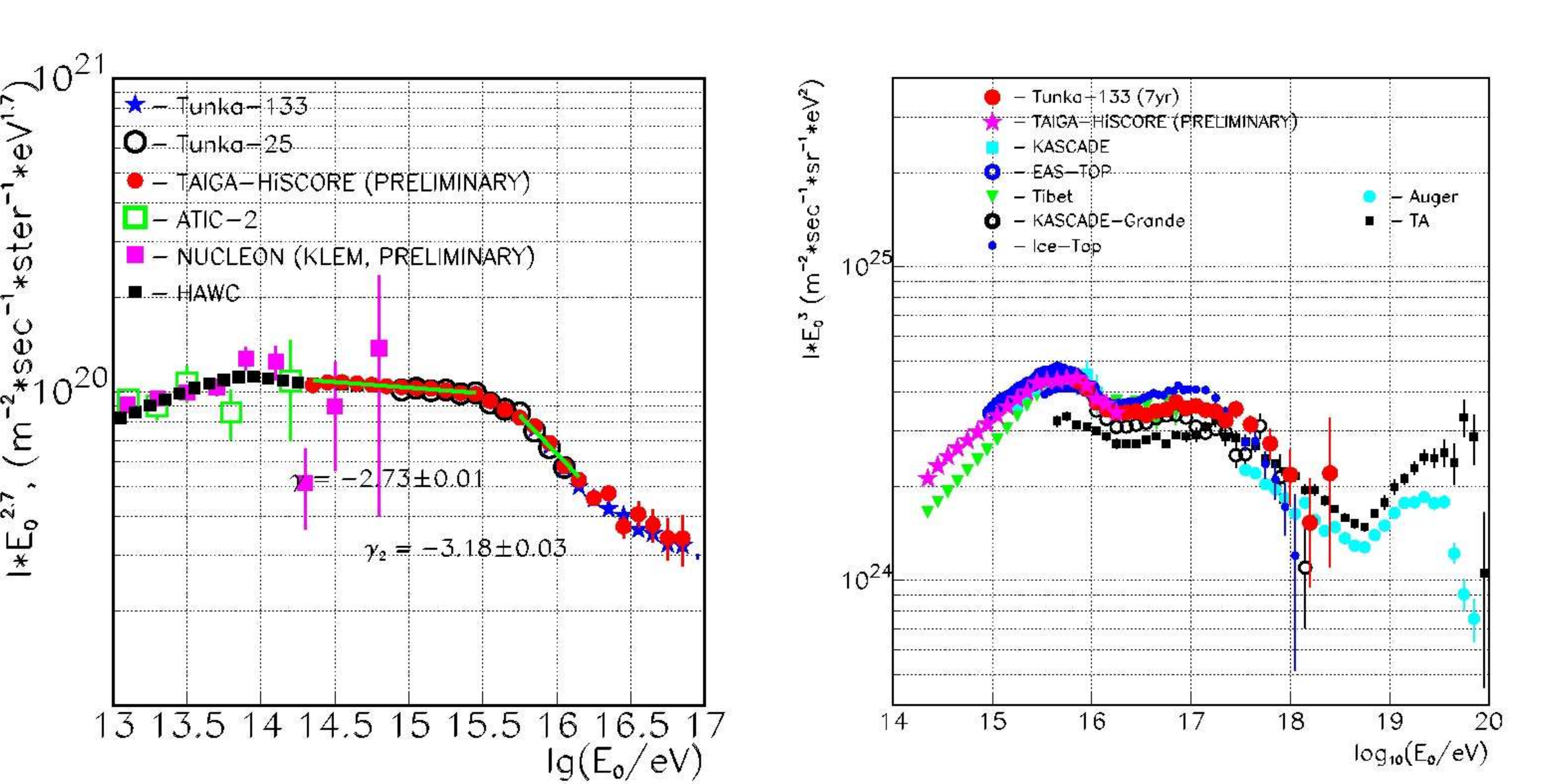}
\caption{
\textit{Top:} The flux of cosmic rays in energy range from the knee to the galactic-extragalactic transition reconstructed by Tunka-25 and Tunka-133 setups (zoomed in right panel).
\textit{Bottom:} Cosmic ray spectrum extended with measurements of TAIGA-HiSCORE in the low-energy region compared to balloon and satellite experiments and ground setups~\cite{Prosin_ECRS2018}.
}
\label{fig:energy_spectrum}
\end{figure}
The cosmic-ray setups of TAIGA are optimized for measurements in the energy range up to a few EeV.
The main features of this region are the first and second knee and the transition from galactic to extragalactic accelerators of cosmic rays.
The first feature has been studied by measuring the absolute flux of the cosmic rays (see Fig.~\ref{fig:energy_spectrum}).
Since Tunka-133 is operating in the same energy range as KASCADE-Grande~\cite{Apel:2013uni}, it provides a cross-check of the position of second knee using a model-independent reconstruction with air-Cherenkov techniques~\cite{Prosin:2016rqu}.
By using the Tunka-Rex and LOPES radio extensions of Tunka-133 and KASCADE-Grande, respectively, it was shown that both detectors feature a consistent energy scale~\cite{Apel:2016gws}, what confirmed the determination of the spectral shape in this energy region.

\begin{figure}
\centering
\includegraphics[height=6cm]{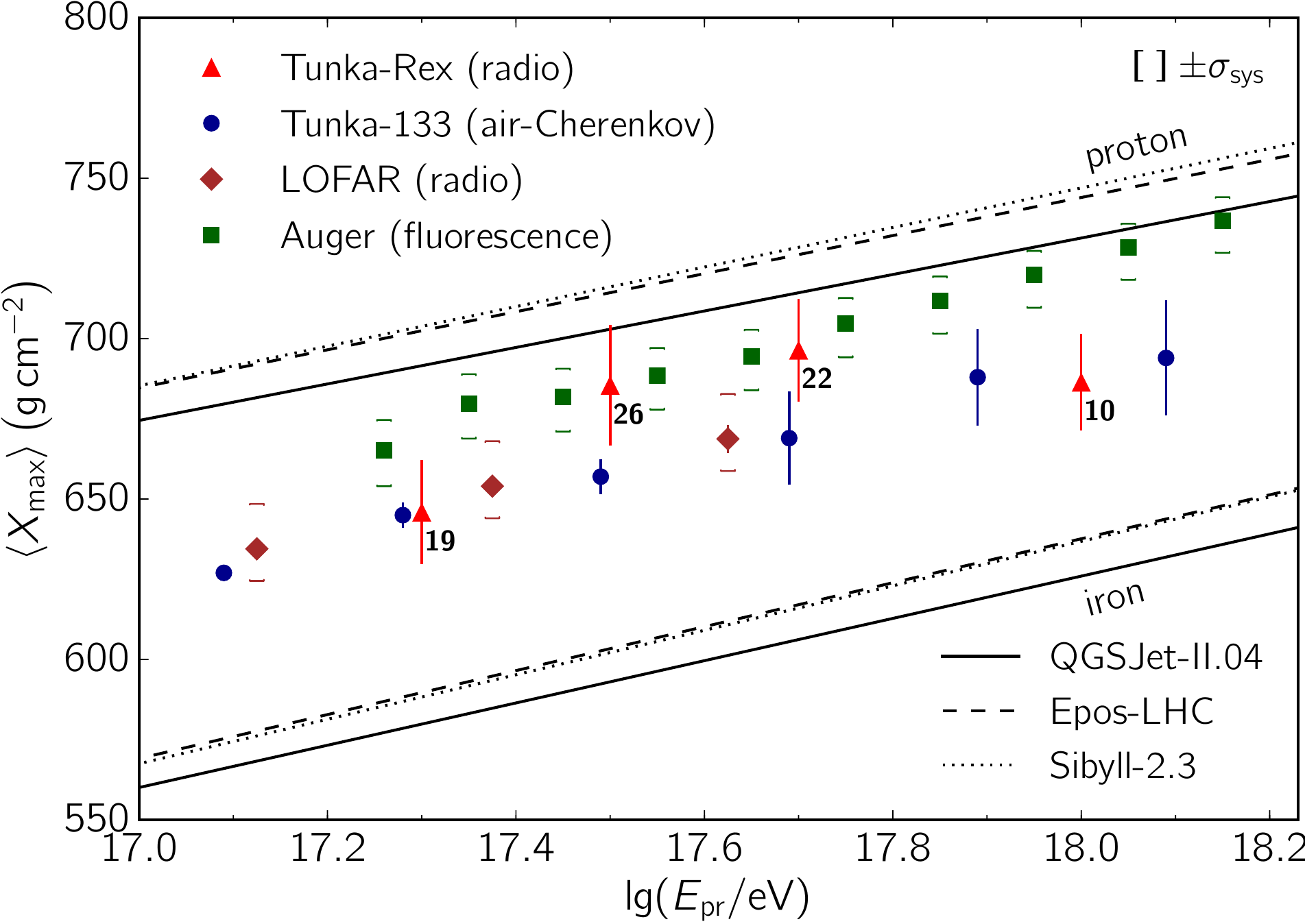}~~~~~~\includegraphics[height=6cm]{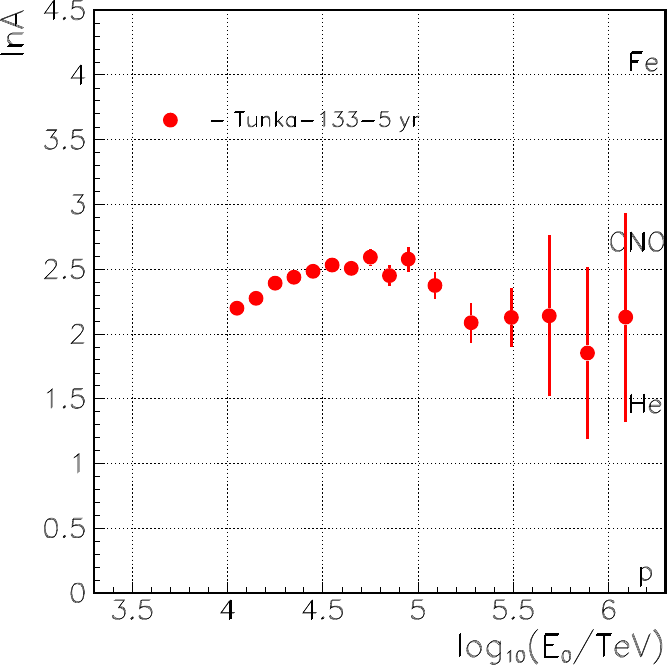}
\caption{\textit{Left:} Mean atmospheric depth of the shower maximum as a function of the energy reconstructed by experiments measuring the electromagnetic component of air-showers~\cite{Prosin:2016jev,Bezyazeekov:2018yjw,Buitink:2016nkf,Bellido:2017cgf}, the model curves are from Refs.~\cite{Ostapchenko:2010vb,Pierog:2006qv,Riehn:2015oba}. 
\textit{Right:} mean mass of primary cosmic rays reconstructed by Tunka-133}.
\label{fig:mass_comp}
\end{figure}

While yielding consistent measurements of the cosmic-ray flux, the instruments from Tunka facility could not add more clarity to the mass composition in the region of the transition between galactic and extragalactic components of cosmic rays (energies of 0.1--1~EeV).
While all experiments disfavor heavy components, the fractions of lighter ones is still very uncertain (see Fig.~\ref{fig:mass_comp}).
Future analyses of Tunka-Rex data combined with particle measurements by Tunka-Grande can decrease this uncertainty and shed light on the mass composition in the transition region.

Besides solving primary astroparticle problems, cosmic-ray experiments in the Tunka Valley are contributing to the development of the different techniques for air-shower detection.
The techniques of air-shower detection using a non-imaging air-Cherenkov array was significantly improved and successfully applied to Tunka-133.
The Tunka-Rex detector has developed and tested many successful techniques of data analysis in the relatively young field of air-shower detection with digital radio arrays~\cite{Schroder:2016hrv}.
Finally, with the recently commissioned Tunka-Grande array the new technique of combined analysis of secondary particles (electrons and muons) and electromagnetic energy (Cherenkov light and radio) will be implemented and validated.

\begin{figure}[t]
\includegraphics[width=1.0\linewidth]{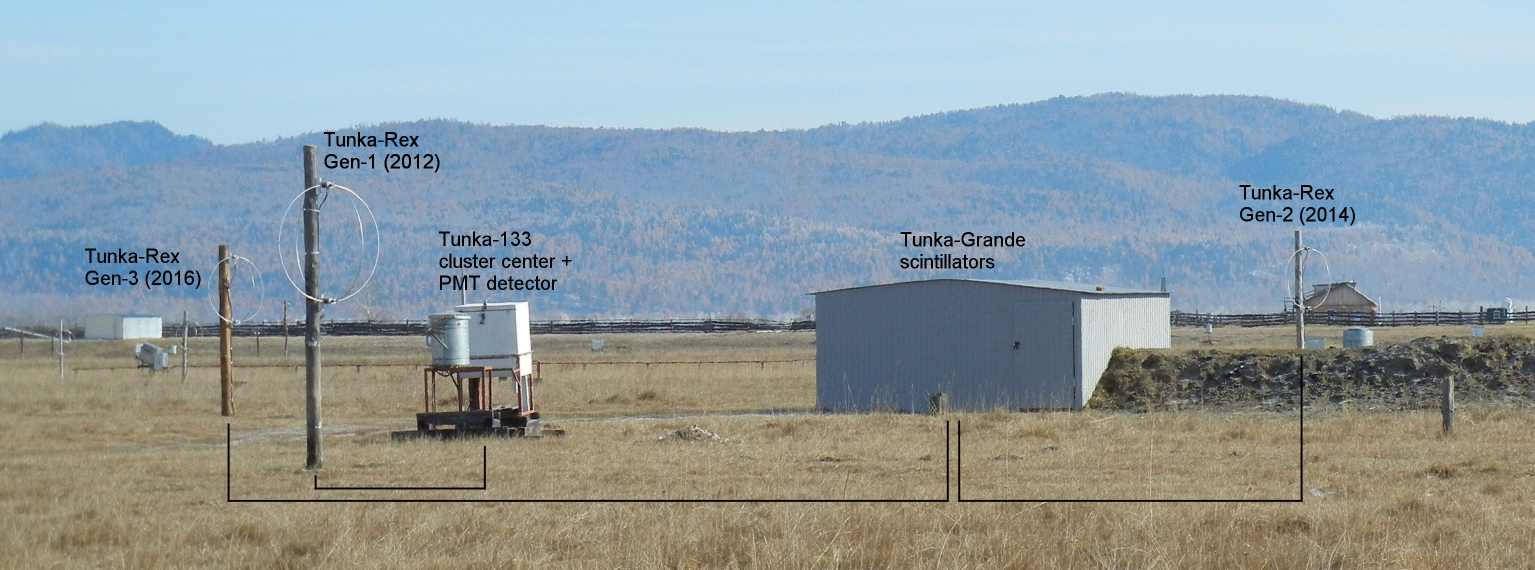}
\caption{Photo of a single cosmic-ray cluster of the TAIGA facility. Lines mark the cable connections between Tunka-Rex antennas and the DAQ of Tunka-133 and Tunka-Grande.}
\label{fig:tunka-cluster}
\end{figure}

\begin{figure}[t]
\includegraphics[width=1.0\linewidth]{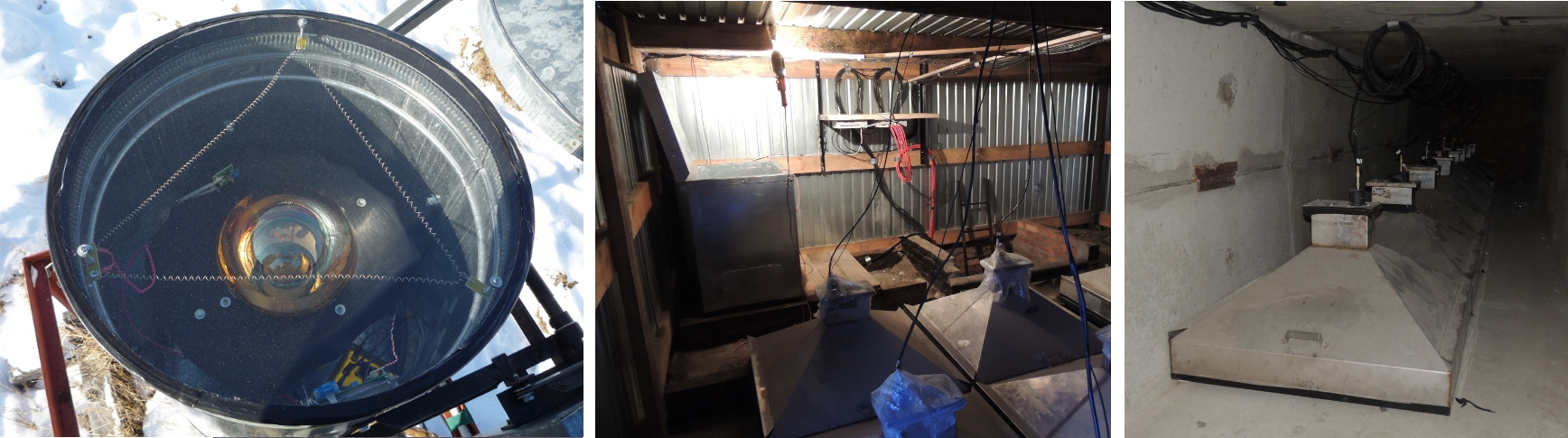}
\caption{Photo of the interior of a Tunka-133 optical module (\textit{left}), on-ground (\textit{center}) and underground (\textit{right}) Tunka-Grande scintillators.}
\label{fig:tunka-interior}
\end{figure}

\begin{figure}[t]
\centering
\includegraphics[width=1.0\linewidth]{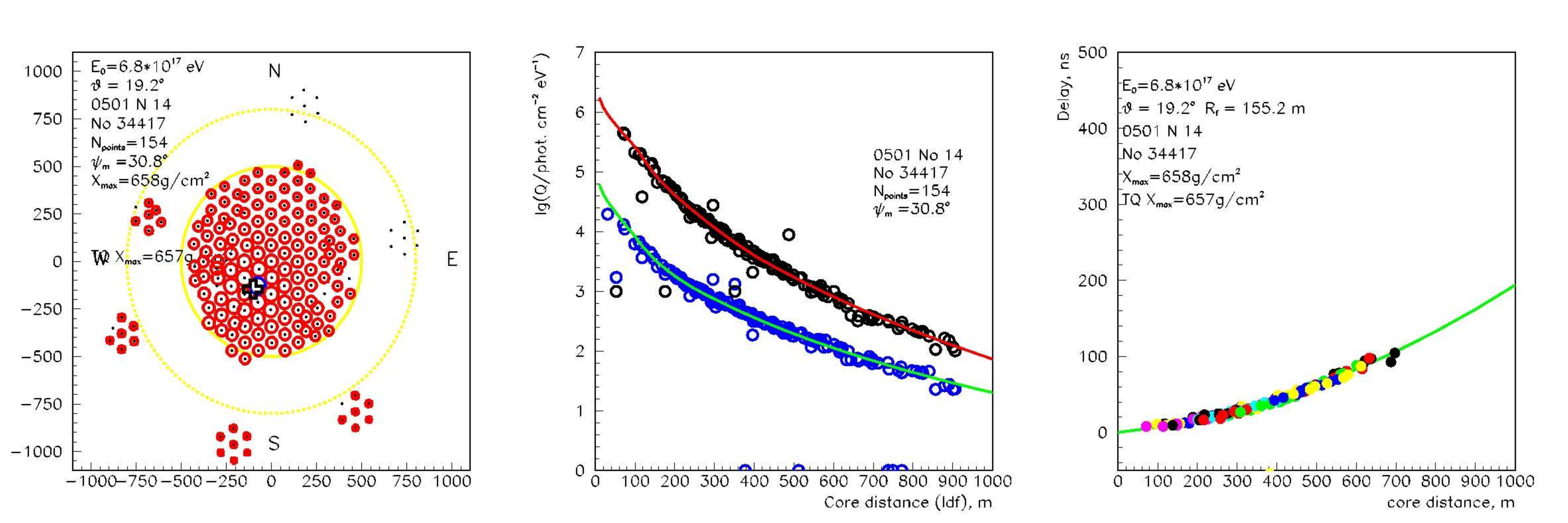}\\
~\\
\hrule
~\\
~\\
\includegraphics[height=3.8cm]{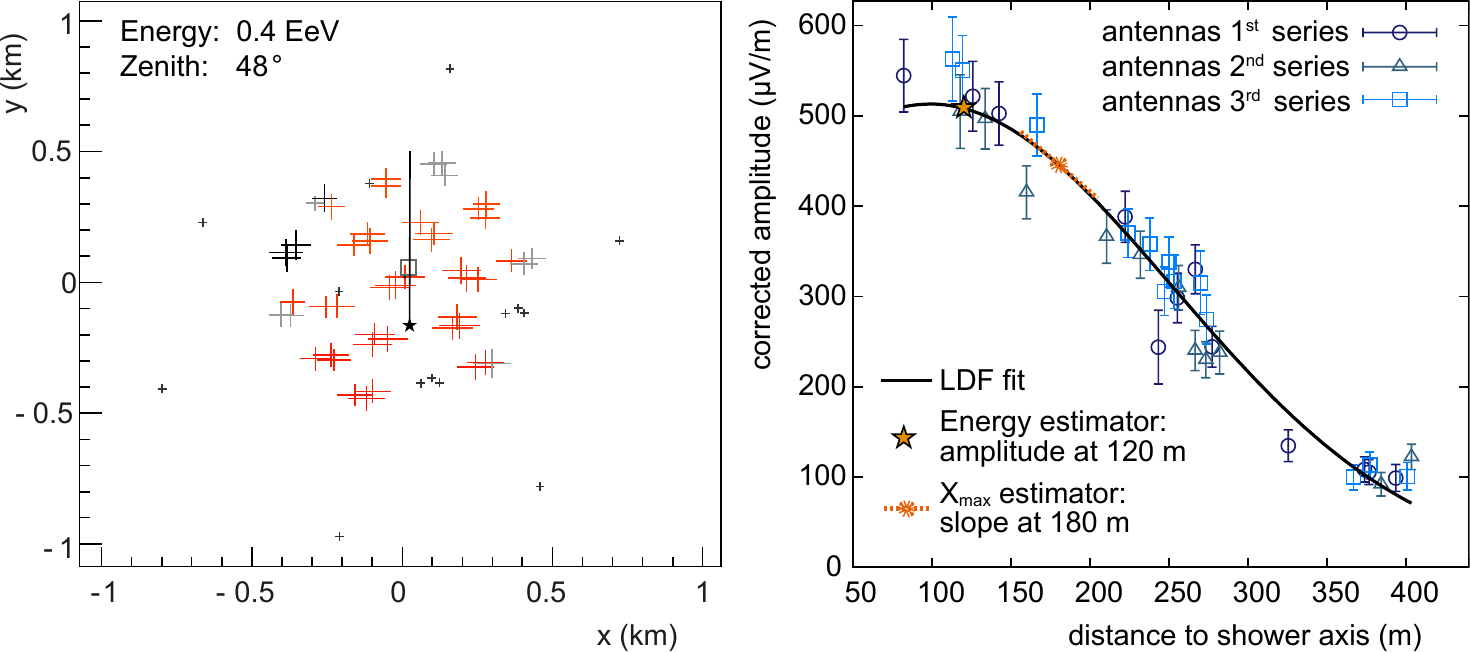} \includegraphics[height=3.8cm]{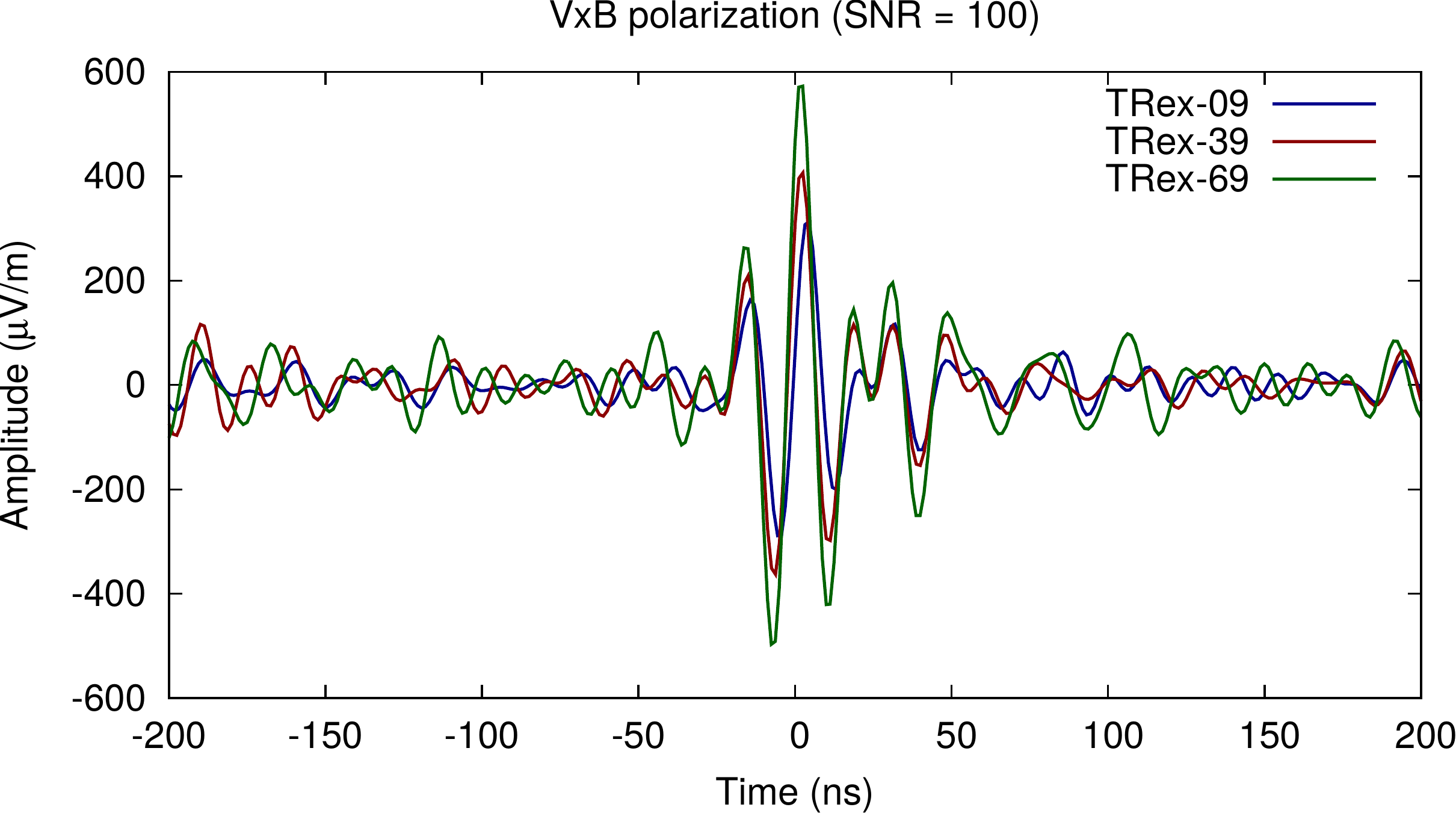}\\
~\\
\hrule
~\\
~\\
\includegraphics[width=1.0\linewidth]{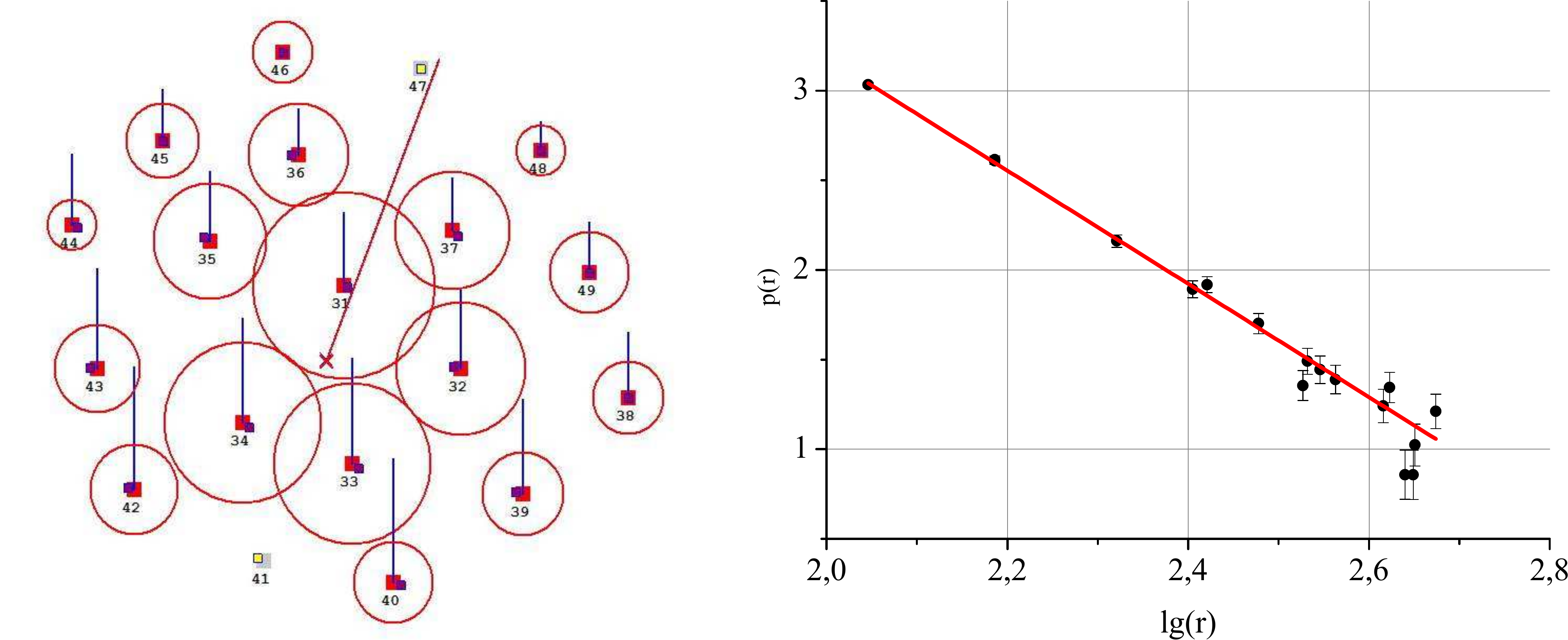}\\
\caption{Example events reconstructed by the Tunka cosmic-ray detectors.
Each detector features its own reconstruction software.
\textit{Top:} Event reconstructed by Tunka-133 (from left to right): 
footprint with triggered detectors -- the cross denotes the shower core, the sizes of circles are proportional to the amplitudes of the optical modules; 
amplitudes and integrals of PMT signals as function of the distance to the shower axis;
the shape of the shower front. 
\textit{Center:} Event reconstructed by Tunka-Rex: footprint, lateral distribution function and signal traces from one of the clusters.
\textit{Bottom:} Footprint and lateral distribution function reconstructed by Tunka-Grande.
}
\label{fig:example-cr-event}
\end{figure}

As the reader can seen in Fig.~\ref{fig:taiga-map}, the cosmic-ray detectors are arranged in 19 clusters over 1~km\textsuperscript{2} (dense core of the setup) and in 6 satellite clusters extending the area to 3~km\textsuperscript{2}.
The photo of a single Tunka cluster from the dense area is shown in Fig.~\ref{fig:tunka-cluster}.
The interiors of a Tunka-133 optical module and Tunka-Grande detectors are given in Fig.~\ref{fig:tunka-interior}.
All three setups are equipped with similar data acquisition (DAQ) electronics which 12 bit-sampling at a rate of 200 MHz; the data are collected in traces made of 1024 samples each.
Each detecting element (Tunka-133 optical module, Tunka-Grande scintillators or Tunka-Rex antenna) has two channels, which are connected via coaxial cables to the local DAQ.
Each cluster is triggered independently, the signals from the active detectors are digitalized and sent to the central DAQ via optical fibers, these optical cables also conduct time synchronization between clusters of about 5~ns.
During clean moonless nights the cosmic-ray setup operates in a triplex mode (air-Cherenkov light, particles and radio) triggered by Tunka-133, the rest of the time it is triggered by Tunka-Grande recording only particle and radio traces.
It is worth noting, that each detector is analyzed with its own unique analysis pipeline, what gives one an opportunity of independent cross-checks on the same events.
Examples of reconstruction from all three detectors are given in Fig.~\ref{fig:example-cr-event}.
The operation period of TAIGA setups usually lasts from October to May, which is motivated by the short dark time during summer time, maintenance activity and lack of lightning protection.
The detailed descriptions of each detector and its features are given below in this section. 

\textbf{Tunka-133}, the pioneer of large-scale setups in the valley, consists of 165 optical modules (OM) grouped in 25 clusters. Each cluster consists of 7 OMs arranged in a hexagon (one OM is placed in the center of the cluster), with a distance of about 80~m between the OMs.
Each optical module consists of a 20~cm PMT housed by a 50~cm metallic cylinder~\cite{Lubsandorzhiev:2014asa}.
Due to its dense layout Tunka-133 features an energy resolution of about 10\%, a shower maximum resolution of about 25~g/cm\textsuperscript{2} and a core resolution of about 10~m for high-energy events~\cite{Prosin:2016jev}.

\textbf{Tunka-Rex} in its final stage consists of 57 antenna stations located in the dense core (3 per cluster) and 6 satellite antenna stations (1 per cluster).
Each Tunka-Rex antenna station consists of two perpendicular active Short Aperiodic Loaded Loop Antennas (SALLA)~\cite{Abreu:2012pi}, the signals are pre-amplified with a Low Noise Amplifier (LNA).
Signals from antenna arcs are transmitted via 30~m coaxial cables to the analog filter-amplifier, which passfilters a frequency band of 30-80~MHz, and later are digitalized by the local DAQ.
One of the goals of the Tunka-Rex detector was the development and test of techniques for the precise reconstruction of air-showers. This goal was successfully achieved: analytical and Monte-Carlo-driven methods were developed and semi-blindly cross-checked with Tunka-133 reconstruction~\cite{Kostunin:2015taa,Bezyazeekov:2015ica,Bezyazeekov:2018yjw}.
Besides this, Tunka-Rex actively uses machine learning techniques~\cite{Bezyazeekov:2017jng,Shipilov:2018wph} and develops methods for reconstruction of inclined events~\cite{Marshalkina:2018sfx}.
Since Tunka-Rex features absolute energy calibration, it (and radio in general) can serve as a good tool for cross-check of systematics in the energy spectrum~\cite{Fedorov:2017xih, Lenok:2018das}.

\textbf{Tunka-Grande} is a scintillator array consisting of 19 detectors located in the dense core of the Tunka clusters.
Each detector features 8~m\textsuperscript{2} surface and 5~m\textsuperscript{2} underground scintillators constructed from the former KASCADE-Grande scintillators.
Tunka-Grande was put into operation in 2015. Calibration and adjustment are still in progress, however simulations and preliminary analyses show promising results in the reconstruction of cosmic rays, especially in combination with Tunka-133 and Tunka-Rex~\cite{Monkhoev:2017gds, Monkhoev:2017yhj}.

\section{Gamma-ray setups}
\begin{figure}[t]
\centering
\includegraphics[height=6cm]{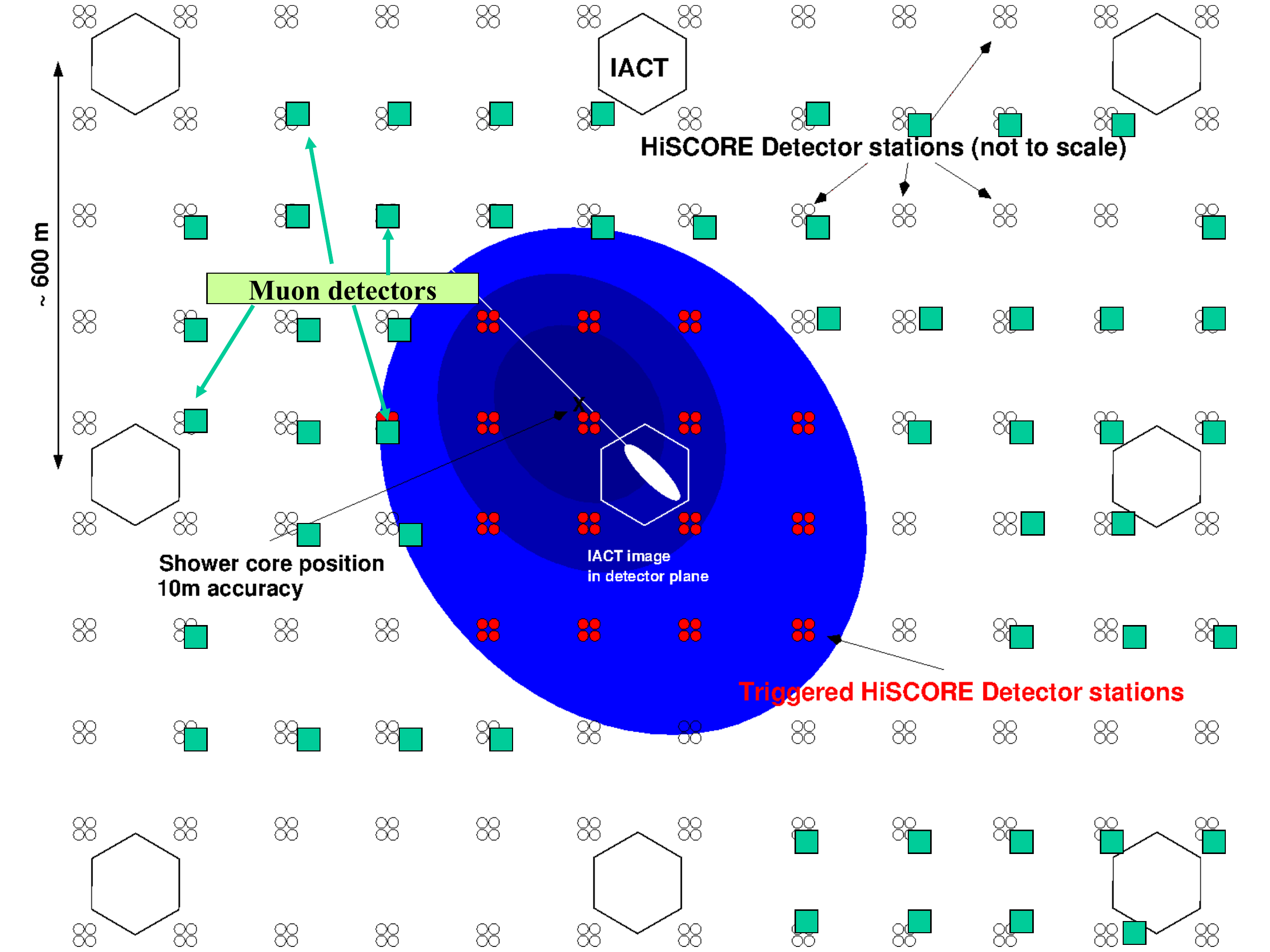}~\includegraphics[height=6cm]{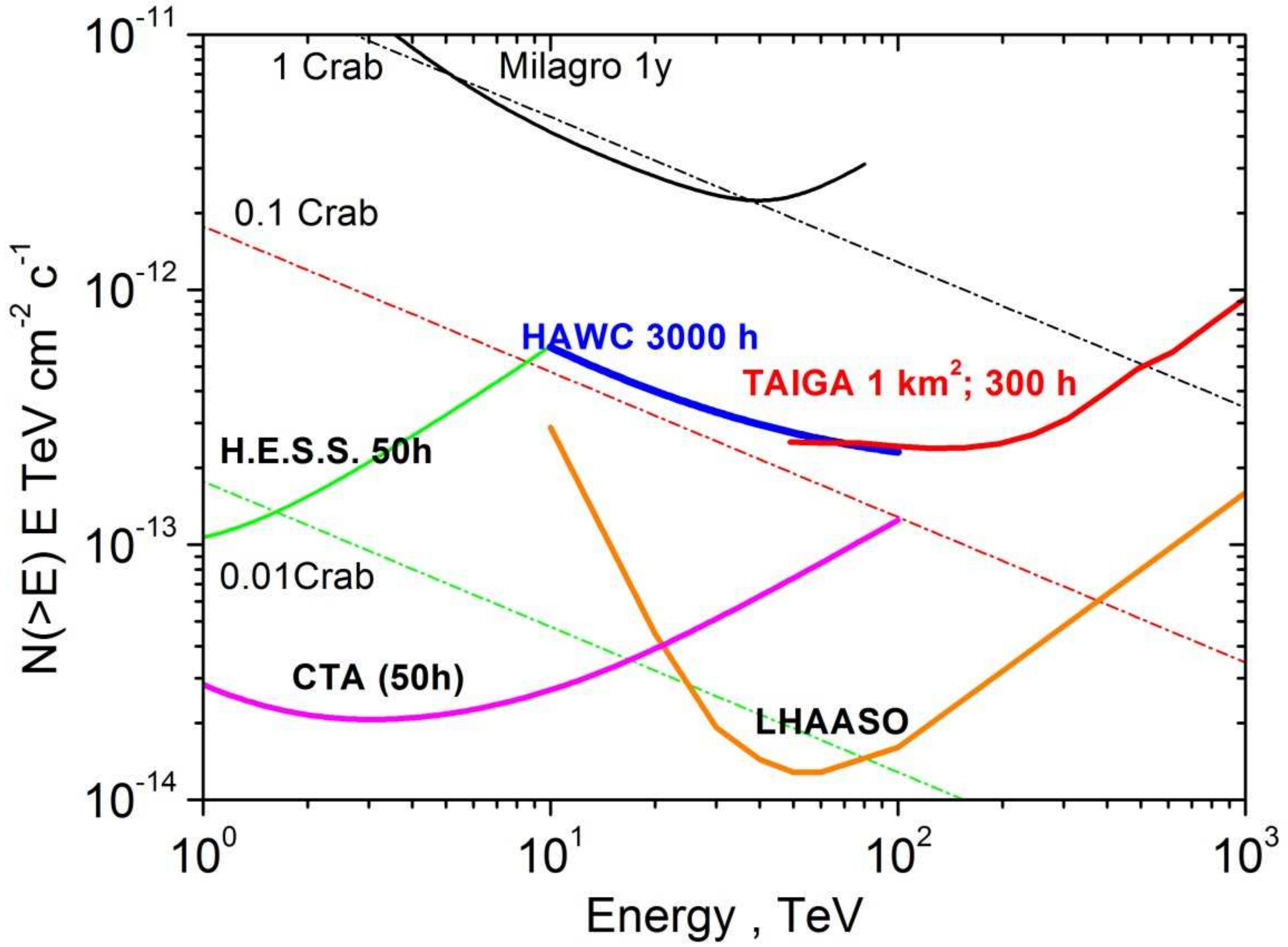}	
\caption{\textit{Left:} To cover very-high energy range it was suggested to apply a hybrid concept for TAIGA.
The wide-angle non-imaging array TAIGA-HiSCORE provides direction and core reconstruction for a net of imaging telescopes TAIGA-IACT, which adds superior gamma-hadron separation.
This configuration allows one to cover a large area with telescopes operating in monoscopic mode.
To improve the quality of gamma-hadron separation and to increase the sensitivity to diffuse gamma rays, an array of underground muon detectors TAIGA-Muon will be constructed complementary to TAIGA-HiSCORE.
\textit{Right:} sensitivity of TAIGA compared to other existing and planned gamma-ray experiments.
}
\label{fig:taiga-concept}
\end{figure}

TAIGA has a unique concept of gamma-ray detection, which combines both imaging and non-imaging techniques.
The wide-angle non-imaging air-Cherenkov array TAIGA-HiSCORE allows one to reconstruct energy, arrival direction and position of the core covering zenith angles up to 60$^\circ$.
At the same time, a sparse net of Imaging Atmospheric Cherenkov Telescopes (IACT) TAIGA-IACT operating in monoscopic mode provides high-efficiency gamma-hadron separation using the geometry reconstruction from TAIGA-HiSCORE.
This configuration allows TAIGA to push the sensitivity to gamma rays towards highest energies ($>50$~TeV) and makes it complementary to existing imaging and non-imaging telescopes.
Moreover, TAIGA is the most Northern telescope (longitude of $\approx52^\circ$), which opens prospects of studying unexplored regions of the TeV sky.
One can see the concept of TAIGA and its sensitivity in Fig.~\ref{fig:taiga-concept}.

\begin{figure}[t]
\centering
\includegraphics[height=5cm]{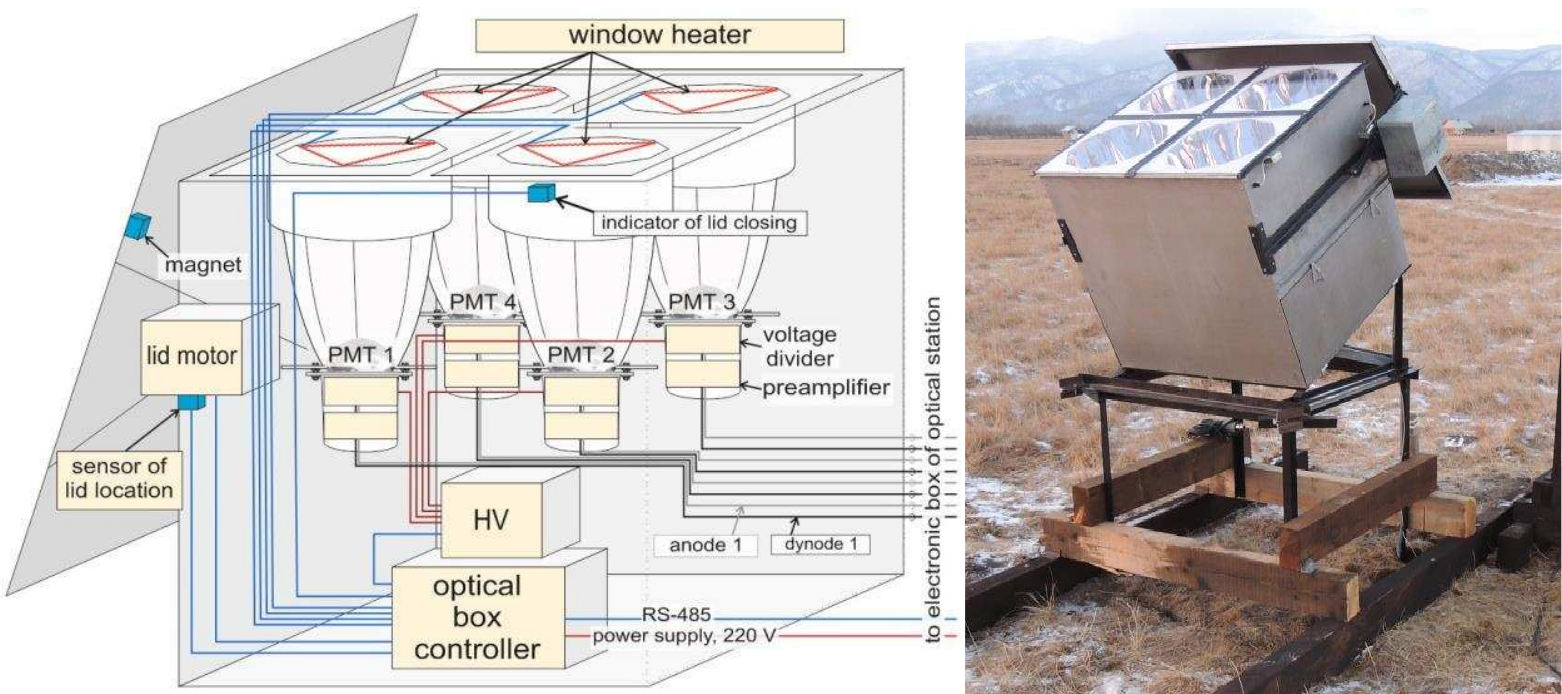}~\includegraphics[height=5cm]{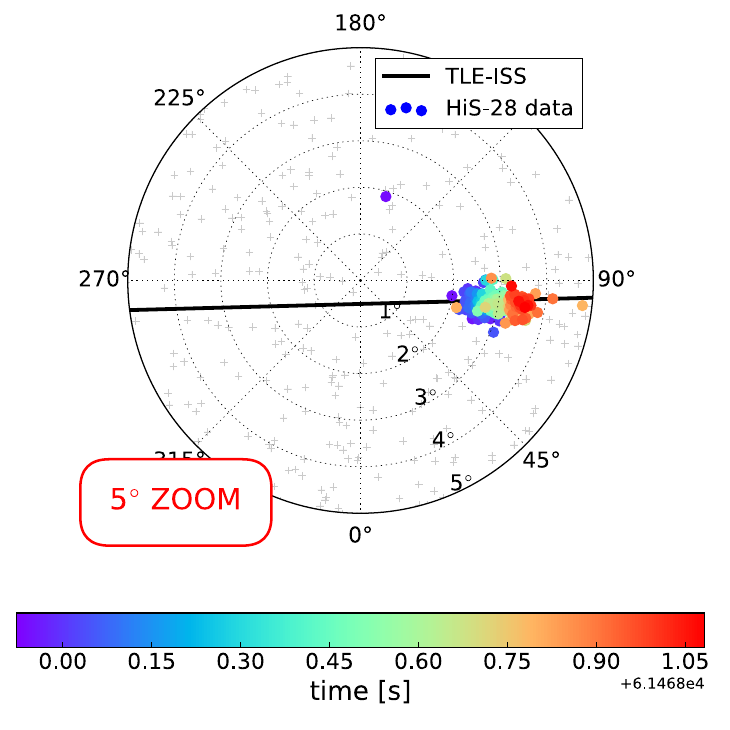}
\caption{\textit{Left and center:}~TAIGA-HiSCORE design and photo of a deployed station. 
\textit{Right:}~Skymap footprint of the ISS lidar detected by TAIGA-HiSCORE.}
\label{fig:hiscore}
\end{figure}

At the moment TAIGA-HiSCORE consists of about 60 stations distributed over an area of 0.6~km\textsuperscript{2}.
Each station is equipped with four 8-inch PMTs with a total light collection area of 0.5~m\textsuperscript{2} and a 0.6 sr field of view.
The data acquisition of TAIGA-HiSCORE differs from the one used in the Tunka cosmic-ray setups and are based on DRS boards with 2~GHz sampling rate.
The array operates with 20 Hz count rate, the trigger is based on the sum of the signals from all four PMTs, which are digitalized locally and sent to the central DAQ via optic fibers.
Precise timing synchronization allows sub-ns reconstruction of air-shower arrival times, what is important for the high angular resolution~\cite{Tluczykont:2017led}.
One can see the design of a TAIGA-HiSCORE station in the left panel of Fig.~\ref{fig:hiscore}.

\begin{figure}[t]
\centering
\includegraphics[width=1.0\linewidth]{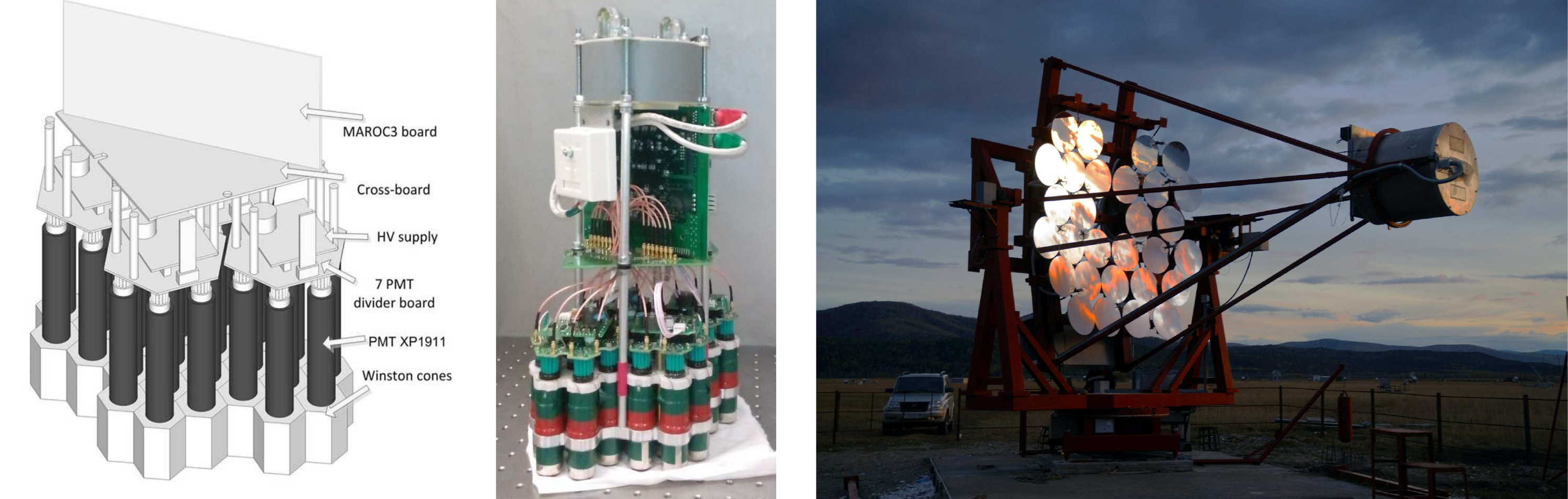}
\caption{\textit{Left and center:}~Design and photo of a TAIGA-IACT camera cluster, which contains 28 PMTs and an independent board generating the trigger.
\textit{Right:}~Photo of the first TAIGA-IACT telescope with 29 mirror segments installed.}
\label{fig:iact-camera-photo}
\end{figure}

\begin{figure}[t]
\centering
\includegraphics[width=1.0\linewidth]{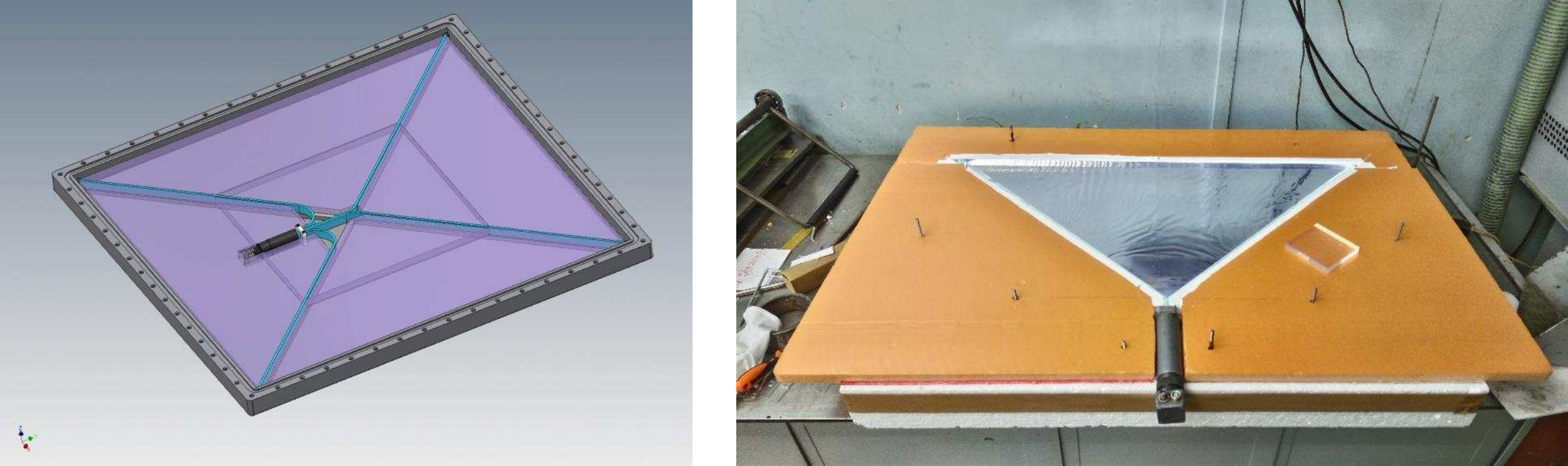}
\caption{\textit{Left:}~The design of the TAIGA-Muon counter.
\textit{Right:}~Photo of a prototype TAIGA-Muon counter.}
\label{fig:taiga-muon}
\end{figure}

The TAIGA-HiSCORE stations are slightly inclined in order to increase their sensitivity to the Crab nebula.
Monte Carlo studies cross-checked with measurements and the comparison against Tunka-133 have shown that the array features an angular resolution up to 0.1$^\circ$ and an energy resolution of up to 10\%.
An additional cross-check has been done when TAIGA-HiSCORE detected the lidar installed on the ISS~\cite{Porelli:2017blz}.
The signal appeared in the array as a plane-wave source moving with a velocity of 7 km/s. One can see its skymap footprint in the right panel of Fig.~\ref{fig:hiscore}.

The first TAIGA-IACT telescope was put into operation in 2017.
Its mechanical part has a HEGRA-like design~\cite{Puhlhofer:2003ir} with a 4.3~m Davies-Cotton mirror consisting of 34 segments with diameter 60~cm, the focal length of telescope is 4.75~m.
The camera comprises 560 hexagonal-shaped pixels with 19~mm XP1911 PMTs equipped with Winston cones.
The PMTs had formerly worked in the calorimeter of the ZEUS experiment in DESY, Hamburg, and have been delivered by DESY and by Gent University.
The field of view of a single pixel is 0.36$^\circ$, which results in a full field of view of 9.72$^\circ$~\cite{Lubsandorzhiev:2017tby}.
One can see the design and photo of telescope and camera in Fig.~\ref{fig:iact-camera-photo}.

The pointing of the telescope is calibrated with a CCD-camera Prosilica GC1380, which is installed at a distance of 1 m from the telescope optical axis on the dish near the mirrors.
The CCD camera observes the same sky as the camera of the telescope and allows on-fly correction of the tracking.
First measurements have shown that the precision of the tracking is better than $0.05^\circ$~\cite{Zhurov_ECRS2018}.

In the winter 2017-2018, TAIGA-IACT has performed about 25~h of measurements targeting the Crab nebula and Mkr421. The data is under analysis~\cite{Sveshnikova_ECRS2018}.

Complementary to optical instruments, there is TAIGA-Muon, a planned array of underground muon detectors, aimed at improving gamma-hadron separation.
The detector will be based on 1~m\textsuperscript{2} cost-effective mass-produced scintillators based on polystyrene with thickness 10-20 mm equipped with 25-46 mm PMTs.
One can see the design and prototype photo of TAIGA-Muon in Fig.~\ref{fig:taiga-muon}.

\section{Open data and outreach}
Modern astroparticle science develops with increasing complexity of instruments, methods, software, data formats, etc.
Combination of these factors led the scientific community to the problem of data life cycle, which resides in data description, conservation, refining and reusing.
There are a number of different approaches to this problem, some of them are suggested in the frame of the German-Russian Astroparticle Data Life Cycle Initiative (GRADLCI)~\cite{Bychkov:2018zre}.
As a partner of this initiative, TAIGA provides a platform for testing algorithms of deep learning on simulated and real data.
The system of data storage and collaboration use is actively developing and will be tested on TAIGA and KCDC~\cite{Haungs:2018xpw} data with the future possibility of combined analyses of measurements from several experiments.
Last, but not least, the educational and outreach programs will be developed and released on the \texttt{astroparticle.online} platform, which will consist of different features, e.g. access to open astroparticle data and software, online courses and tutorials.

\begin{figure}[t]
\centering
\includegraphics[width=1.0\textwidth]{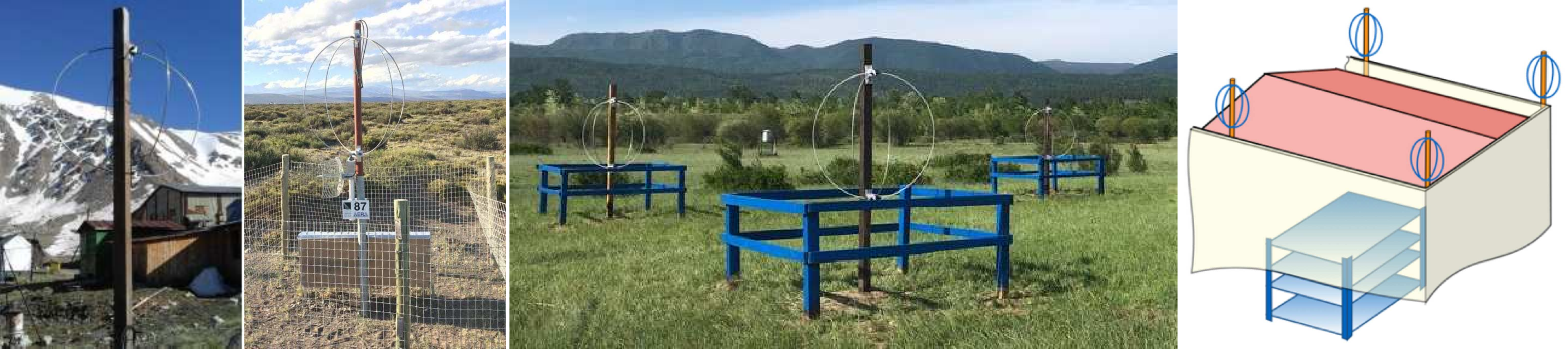}
\caption{Tunka-Rex SALLA installed at different sites. \textit{From left to right:} Tien-Shan High Mountain Cosmic Station (Kazakhstan), Pierre Auger Observatory (Argentina)~\cite{Hoerandel_ARENA2018}, ISU educational cluster (later Tunka-21cm), future TRASGO detector (Spain)~\cite{Belver:2012zz}.}
\label{fig:tunkarex_collab}
\end{figure}

As was mentioned above, the modern technique for detection of ultra-high energy air-showers with digital radio arrays is actively developing at TAIGA in the frame of the Tunka-Rex experiment.
Tunka-Rex has shown the feasibility of its robust and cost-effective antenna (SALLA) and shared technology and software with other cosmic-ray experiments.
Particularly, one more upgrade of the Pierre Auger Observatory was approved~\cite{Hoerandel_ARENA2018}, in frame of which every surface detector will be equipped with SALLA hardware. 
Tunka-Rex antennas and experience have also been used for deployment of educational and engineering arrays in Spain~\cite{Belver:2012zz} and Kazakhstan (high-altitude array Almarac, an Almaty RAdio Cluster).
One can see the photos of these instruments in Fig.~\ref{fig:tunkarex_collab}.

\section{Cooperation with other experiments in the valley}
There are a few neighboring instruments installed on the same site as TAIGA and sharing the infrastructure and resources.
The first is a telescope of the MASTER robotic network~\cite{Lipunov:2009ck}, which performs wide-angle and fast follow-up observations of optical transients, particularly gamma-ray bursts~\cite{Lipunov:2017dwd}.
The second is the geophysical observatory consisting of low-frequency MHz antennas, which belongs to the Institute of Solar-Terrestrial Physics and is mainly aimed at the study of the ionosphere.
The latest one is the recently deployed engineering array Tunka-21cm, a side project of the Tunka-Rex experiment, which aims at the development of methods for detection of the cosmological signal from neutral hydrogen.

\section{Conclusion}
We have overviewed the instruments installed in the Tunka Valley at the TAIGA observatory focused on very- and ultra-high energy cosmic rays and gamma radiation.
A comprehensive description of every detector is given as well as their performance and resolution.
Let us summarize the main achievements of the observatory:
\begin{itemize}
\item Precise study of the cosmic-ray flux in the energy region of $10^{14.5}$--$10^{18.5}$~eV, studying the knees of the spectrum, providing a cross-check with KASCADE-Grande results.
\item Study the region of the transition from galactic to extragalactic accelerators with different techniques (air-Cherenkov, radio).
\item Developing and improving techniques for the measurement of ultra-high energy air-showers with non-imaging arrays using air-Cherenkov, radio and scintillator techniques.
\end{itemize}
Since the observatory develops towards very-high energy gamma astronomy, the plan for the near future (before the end of 2019) is the deployment of about 120 HiSCORE detectors, 3 IACTs and 250~m\textsuperscript{2} of TAIGA-Muon scintillators.
This pilot array will cover about 1~km\textsuperscript{2}.
The full-scale TAIGA observatory is intended to consist of about 1000 HISCORE detectors and 15 IACTs on an area of 10~km\textsuperscript{2}~\cite{Budnev:2018fxf}.
Its unique layout and the hybrid techniques allow TAIGA to measure the highest tails of the spectrum of brightest galactic and extragalactic sources.
The geographical location allows TAIGA to observe unexplored parts of the high-energy gamma-ray sky, particularly the remnant of the Tycho Brahe supernova.

Being the most advanced astrophysical facility in Russia, TAIGA affiliates about 80 scientists from 15 institutions in Russia and Europe (mostly Germany) and conserves very strong and fruitful cooperation with colleagues around the world.

\section*{Acknowledgements} 
The lecturer (Dmitriy Kostunin) thanks the organizers of the ISAPP-Baikal Summer School for the invitation. 
Parts of this work has been supported 
by the Russian Foundation for Basic Research (grants No. 16-29-13035, 17-02-00905, 18-32-00460, 18-32-20220), 
by the Ministry of Science and Higher Education of the Russian Federation (Tunka shared core facilities, unique identifier RFMEFI59317X0005, agreements: 3.9678.2017/8.9, 3.904.2017/4.6, 3.6787.2017/7.8, 3.6790.2017/7.8, 3.6241.2017/6.7, 3.6275.2017/6.7, 3.5917.2017/BCh),
by the Helmholtz grant No. HRSF-0027,
by the European Union's Horizon 2020 programme (ASTERICS No.653477),
and by the grant No. 19-72-20067 of the Russian Science Foundation (Section 4).

\section*{References}
\bibliography{references}
\end{document}